\documentclass[12pt]{article}

\usepackage{amsmath}
\usepackage{amsthm,amssymb}

\usepackage{graphicx}
\numberwithin{equation}{section}

\newcommand{\ma}{m_{as}}
\newcommand{\ms}{m_{ss}}

\newcommand{\ba}{\beta_{as}}
\newcommand{\bs}{\beta_{ss}}

\graphicspath{figs}

\begin{document}

\title{A semi-flexible attracting-segment model of three-dimensional polymer collapse}

\author{J.\ Krawczyk$^{\flat,\dag}$,
A.\ L.\ Owczarek$^\ddag$,
and T.\ Prellberg$^\star$\\[1em]
 \footnotesize
  \begin{minipage}{11cm}
$\flat$ Institute of Molecular Physics, Technical University of {\L}{\'o}d{\'z}, 90-924 {\L}{\'o}d{\'z}, Poland \\[1ex]
$\dag$ Department of Mathematical Sciences, Durham University,
South Road, Durham DH1 3LE, United Kingdom\\[1ex]
$\ddag$ Department of Mathematics and Statistics, The University of Melbourne, 
3010, Australia\\[1ex]
$\star$ School of Mathematical Sciences, Queen Mary University of London,
Mile End Road, London E1 4NS, United Kingdom
\end{minipage}
}
%\pacs{05.70.Fh,87.10.Hk,61.41.+e}

\maketitle
\begin{abstract}
  Recently it has been shown that a two-dimensional model of self-attracting polymers based on attracting segments 
  with the addition of stiffness displays three phases: a swollen phase, a globular, liquid-like phase, and an anisotropic 
  crystal-like phase. Here, we consider the attracting segment model in three dimensions 
  with the addition of stiffness. While we again identify a swollen and two distinct collapsed phases,
  we find that both collapsed phases are anisotropic, so that there is no phase
  in which the polymer resembles a disordered liquid drop. Moreover all the phase transitions are first order.
\end{abstract}

\section{Introduction}

An isolated polymer in solution undergoes a collapse transition from a swollen coil to a collapsed globule as the
temperature is lowered and consequently the quality of the solvent is reduced. The canonical lattice model used to 
describe this scenario is the model of \emph{Interacting Self-Avoiding Walks} (ISAW) on a regular lattice, such as the square or 
simple cubic lattice \cite{orr1947a-a,vanderzande1991a-a} . 

At high temperatures the self-avoiding walk is swollen, in that the fractal dimension of the walk $d_f$ is less than the
fractal dimension of simple random walks. The exponent $\nu=1/d_f$ describes the scaling of the size of the walk, as measured
for example by its radius of gyration, as a function of the length of the walk. It is known that $\nu=3/4$ in two dimensions \cite{nienhuis1982a-a} and $\nu=0.587597(7)$
in three dimensions \cite{clisby2010}.

At low temperatures the self-avoiding walk is collapsed, in that the fractal dimension of the walk $d_f$ is equal to the dimension $d$
of the ambient space. The transition between the swollen coil and the collapsed globule happens at a particular temperature, called the
$\theta$-temperature.

In the standard description of the coil-globule transition, the transition is a tricritical point related to the $N \rightarrow 0$
limit of the $\varphi^4$--$\varphi^6$ O($N$) field theory \cite{gennes1975a-a,stephen1975a-a,duplantier1982a-a}; there is a second-order phase transition with a specific heat exponent
conjectured to be $-1/3$ in two dimensions \cite{duplantier1987a-a} and $0$ in three dimensions with a logarithmic divergence of the specific heat. In two dimensions
the fractal dimension of the walk is expected to be $d_f=7/4$ \cite{duplantier1987a-a} and $d_f=2$ with logarithmic corrections in three dimensions.

The canonical model of interacting self-avoiding walks fits this scenario. In this model one expects the low-temperature state to be a 
liquid drop, i.e., the polymer is compact but disordered.

In stark contrast to this scenario, there is another simple interacting polymer model, the \emph{Interacting Hydrogen-Bond} model (IHB)
\cite{bascle1993a-a,foster2001a-a,krawczyk2007b-:a}, where a pair of sites on the self-avoiding walk acquires a
hydrogen-like bond potential if the sites are (non-consecutive) nearest neighbours, as in the ISAW model, and each site lies on a straight
section of the walk. This model has been introduced in the context of biopolymers where hydrogen bonding plays an important role \cite{pauling1951a-a}.
In contrast to ISAW, this model displays a first-order collapse transition in both two and three dimensions. Here, the low-temperature state is
an anisotropic compact phase described as a polymer crystal.

\begin{figure}[ht!]
  \centering
  \includegraphics[width=0.45\textwidth]{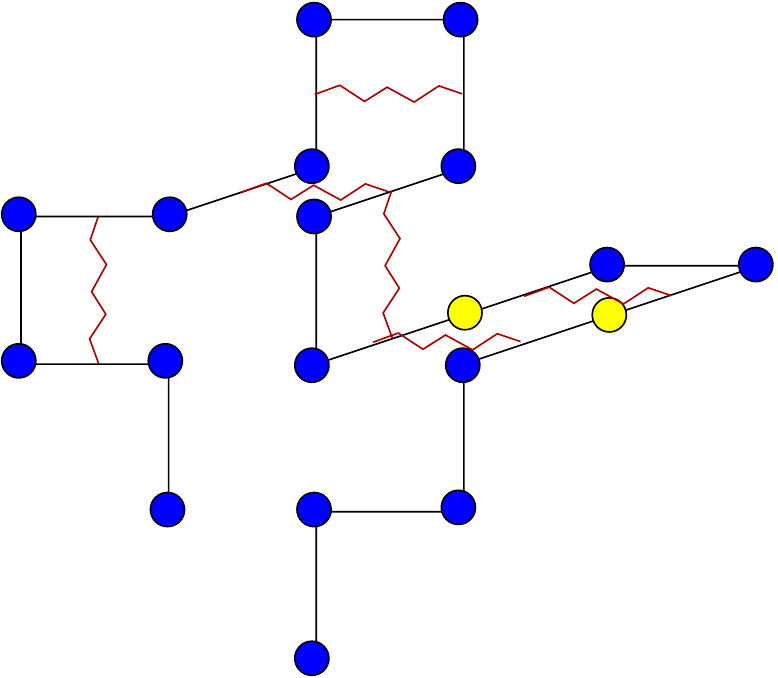}
  \caption{A self-avoiding walk with the interactions of the
    attracting segment (AS) model shown as intertwined curves between
    bonds of the walk on opposite sides of the squares of the
    lattice. Also shown is an example of a stiffness segment pair
    which obtains a stiffness energy in our generalisation.} 
  \label{fig_ssas-model}
\end{figure}

Another model introduced to account for hydrogen bonding is the \emph{Attracting Segments} model (AS) 
\cite{machado2001a-a,buzano2002a-a,foster2007a-a} (also known as `interacting bonds'). It is a lattice model based on self-avoiding walks
where an attractive potential is assigned to \emph{bonds} of the walk that lie adjacent and parallel on the lattice (though not consecutive
along the walk), see Figure~\ref{fig_ssas-model}. On the square lattice, this model seems to have \emph{two} phase transitions, one of which is
identified as the $\theta$-point \cite{foster2007a-a}.

If one introduces stiffness into the ISAW model, one arrives at the \emph{semi-flexible} ISAW model
\cite{bastolla1997a-a,lise1998a-a,vogel2007a-a,doye1997a-a}. In addition to the nearest-neighbour
site interaction of ISAW, one introduces a stiffness energy associated with consecutive parallel bonds
of the walk. This was studied on the cubic lattice by Bastolla and Grassberger
\cite{bastolla1997a-a}, where it was shown that depending on the energetic weighting of straight segments
one finds a single first-order transition from a swollen coil to a crystalline state for a strong energetic
preference for straight segments, or a soft $\theta$-transition from a swollen coil to a liquid globule,
followed by a first-order transition to a crystalline state. In two dimensions, a similar scenario has been found
\cite{krawczyk2009a-:a}, the main difference being that the transition between the globule and the frozen state becomes
second-order.

The semi-flexible AS model, in which both straight segments and interacting segments carry an energy, as shown in Figure \ref{fig_ssas-model},
has been studied in two dimensions in \cite{krawczyk2010a-:a}, where it was found that it has a phase structure in common with the semi-flexible
ISAW model.

In this paper, we discuss the semi-flexible AS model in three dimensions. 
While we again identify a swollen and two distinct collapsed phases,
we find that both collapsed phases are anisotropic, so that there is no phase
in which the polymer resembles a disordered liquid drop. The transitions between the swollen and each of the collapsed phases and between the two collapsed phases are first order. The three lines of transitions in parameter space meet at a triple point.

\section{Our Study}

\subsection{Semi-flexible Attracting Segments model}

Our semi-flexible attracting segments model (semi-flexible AS model)
is a self-avoiding walk on the simple cubic lattice, with
self-interactions as in the AS model
\cite{machado2001a-a,buzano2002a-a,foster2007a-a} and a stiffness (or
equivalently bend energy) added.  Specifically, the energy of a single
chain (walk) consists of two contributions (see
Figure~\ref{fig_ssas-model}): the energy $-\varepsilon_{as}$ for each
attracting segment pair, being a pair of occupied bonds of the lattice
that are adjacent and parallel on the lattice and not consecutive
along the walk; and an energy $-\varepsilon_{ss}$ for each stiffness
segment pair, being a pair of bonds consecutive along the walk that
are parallel. A walk configuration $\varphi_n$ of length $n$ has total
energy
\begin{equation}
E_n(\varphi_n) = -m_{as}(\varphi_n)\ \varepsilon_{as} - m_{ss}(\varphi_n)\ \varepsilon_{ss},
\end{equation}
where $m_{as}$ denotes the number of attracting segment pairs and
$m_{ss}$ denotes the number of stiffness segment pairs.  The partition
function is defined then as
\begin{equation} Z_n(\beta_{as},\beta_{ss})=
\sum_{m_{as},m_{ss}} C_{n,m_{as},m_{ss}} e^{\beta_{as} m_{as}+\beta_{ss} m_{ss}},
\end{equation}
where $\beta_{as}= \varepsilon_{as} /k_BT$ and $\beta_{ss}=
\varepsilon_{ss}/k_BT$ for temperature $T$ and Boltzmann constant
$k_B$.  The density of states, $C_{n,m_{as},m_{ss}}$, 
has been estimated by means of Monte Carlo simulations.

\subsection{Simulations}

On the cubic lattice we performed simulations using the
FlatPERM algorithm \cite{prellberg2004a-a}, estimating the density of
states up lengths for $n=128$ over the two parameters $m_{as}$ and
$m_{ss}$. We use averages for the density of states from five independent runs.
Moreover, we have simulated the density of states in $\ma$ at $\bs=\pm1.0$ and 
in $\ms$ at $\ba=1.0$ for lengths up to $n=256$.

The density of states allows us to calculate the internal energy and the
specific heat, or equivalently, the mean values and the fluctuations
of $m_{as}$ and $m_{ss}$, respectively.
This allows us to locate phase transitions through the possible
divergences in the specific heat.  To detect orientational order, we
estimated an \emph{anisotropy parameter} \cite{bastolla1997a-a}. 

\section{Results}

\subsection{Phase Diagram}

\begin{figure}[ht!]%
\centering
\includegraphics[width=0.45\textwidth]{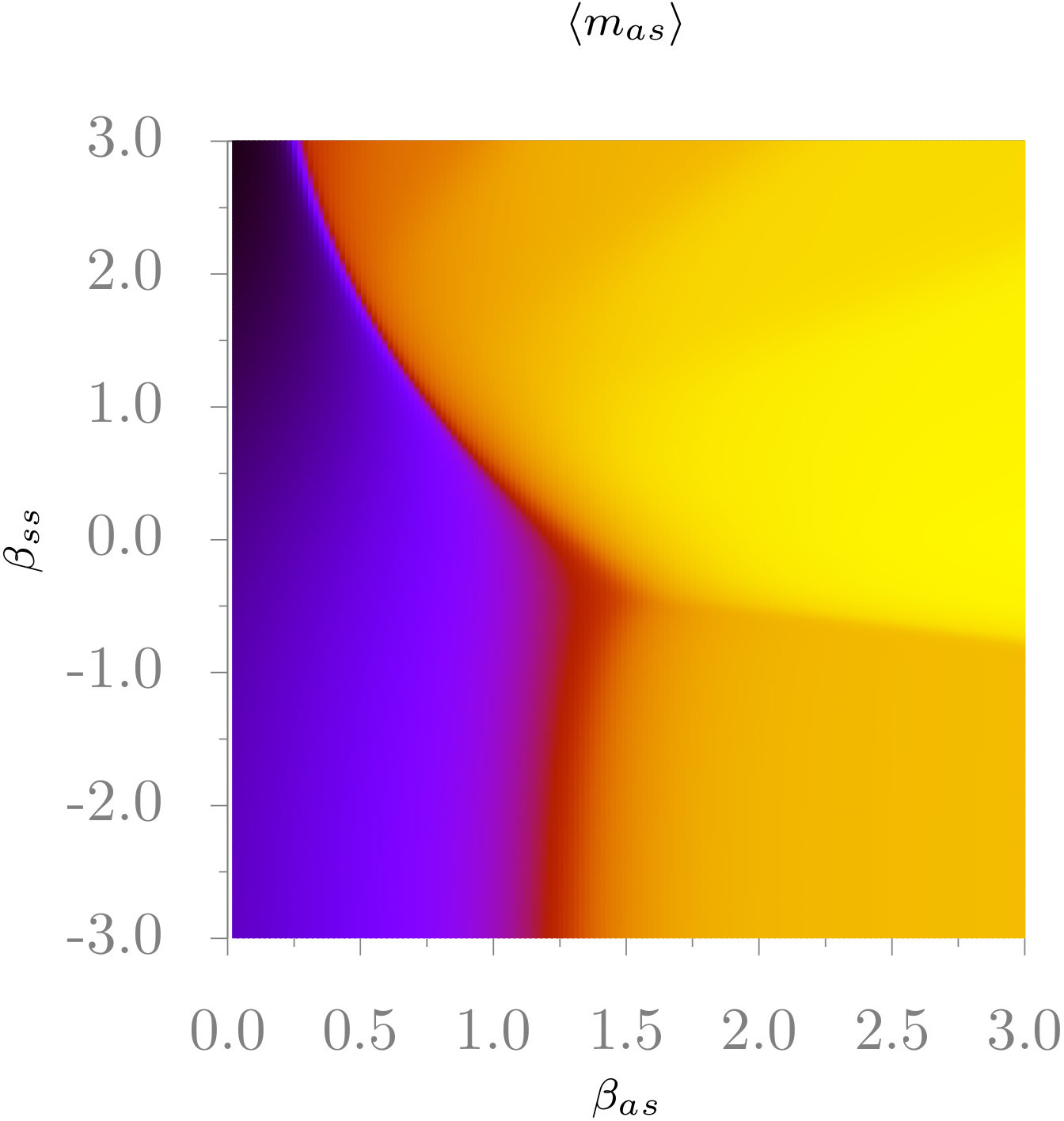}  
\includegraphics[width=0.45\textwidth]{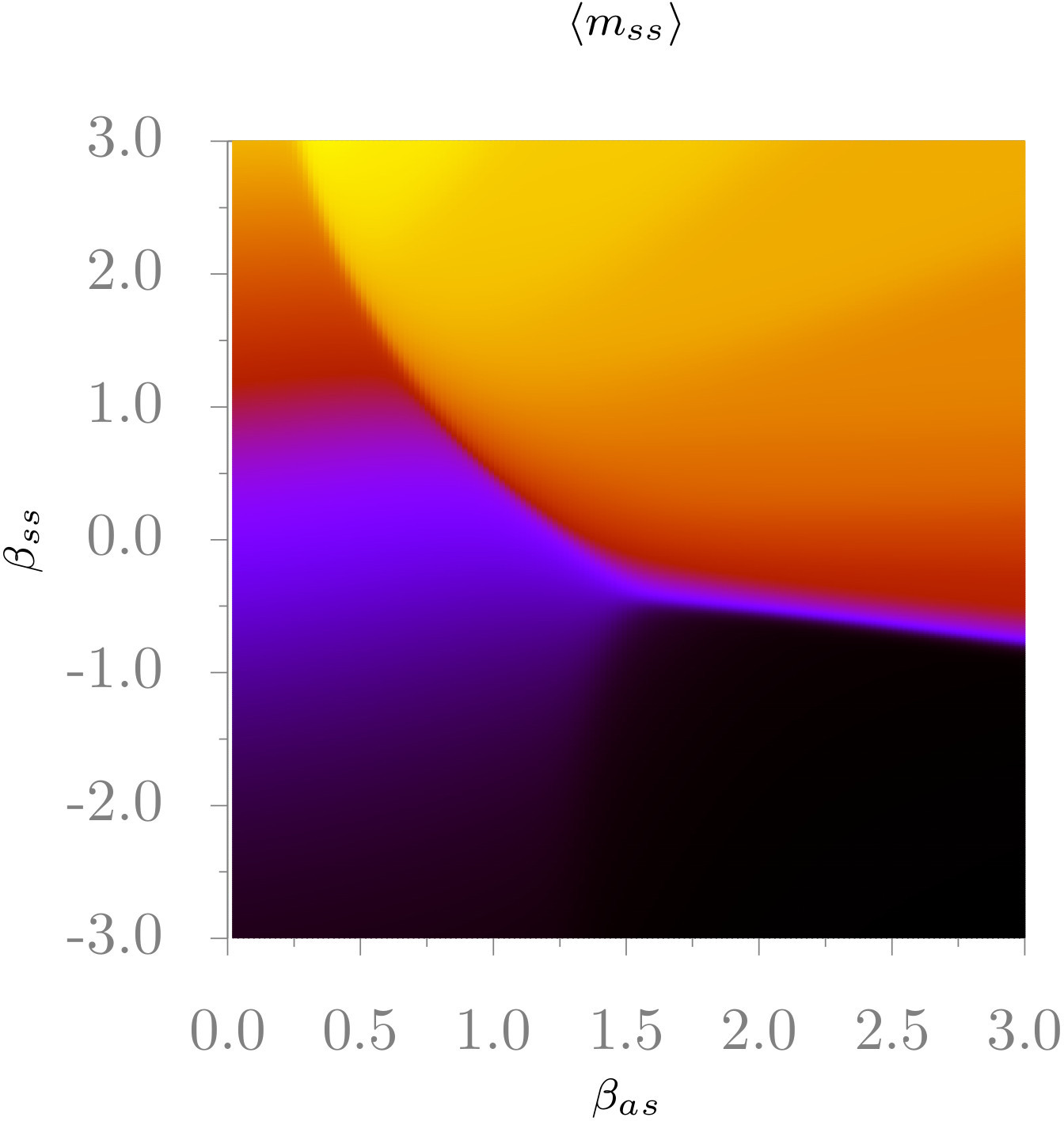}\\
\includegraphics[width=0.45\textwidth]{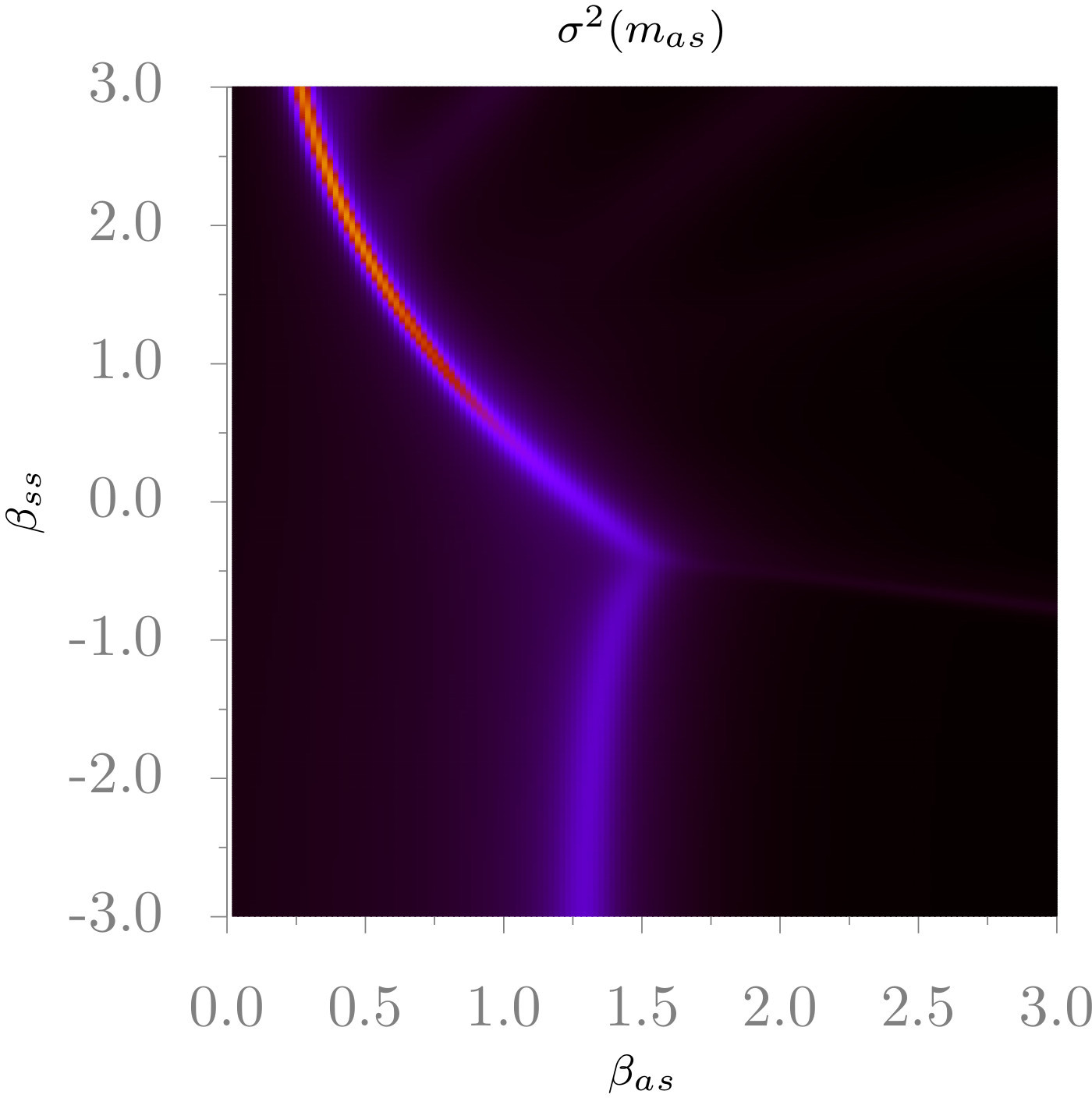}
\includegraphics[width=0.45\textwidth]{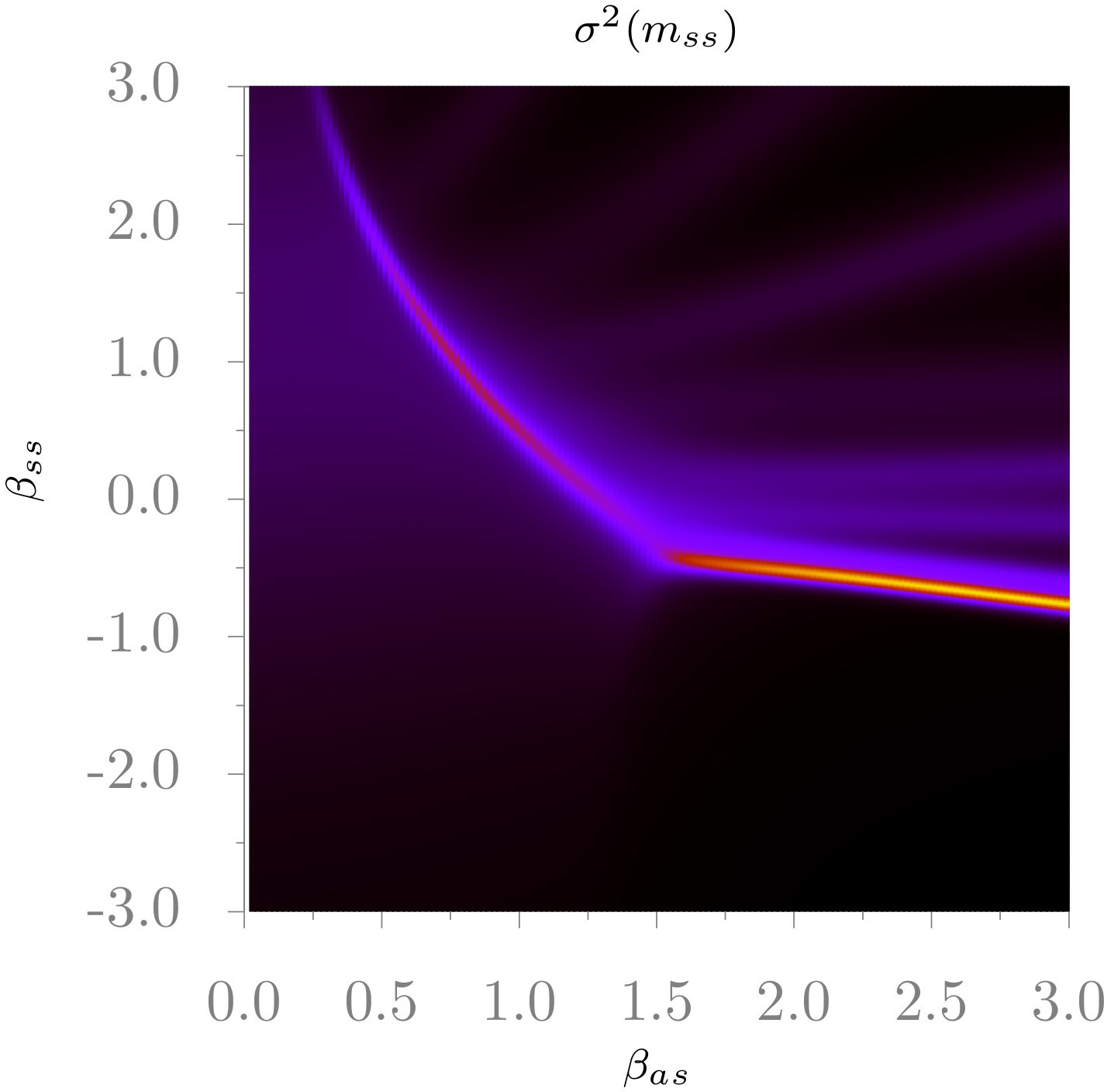}
\caption{\label{s_as} 
Expected value of the number of attracting segments pairs $\ma$ (top left) and stiff segment pairs $\ms$ (top right) and their 
fluctuations $\sigma^2(\ma)$ (bottom left) and $\sigma^2(\ms)$ (bottom right) for $128$-step walks as a function of $\ba$ and $\bs$.}
\end{figure}

From the density of states for $128$-step walks, we compute the expected number of attracting segment pairs and stiff segment pairs, shown in Figure 
\ref{s_as} as a function of $\ba$ and $\bs$. There are three distinct regions recognisable. For small values of $\ba$ both $\langle\ma\rangle$ and $\langle\ms\rangle$ are small, with $\langle\ms\rangle$ increasing slowly as $\bs$ increases. Upon increasing $\ba$, there is a sharp transition to large values 
of $\langle\ma\rangle$, indicative that the walks are collapsed on average. For large $\ba$ the density of straight segments, $\langle\ms\rangle$, is small for negative values of $\bs$, but increases sharply as $\bs$ increases. This indicates that there are two distinct phases when $\ba$ is large (collapsed region).

Putting this broad information together it seems to indicate that there is a region in which the walk forms a swollen coil (the non-interacting SAW at $\ba=\bs=0.0$ lies in this region),
and that there are two regions in which the walk is collapsed. For one of the collapsed regions with negative $\bs$, $\langle\ma\rangle$ is very small (in fact smaller than in the swollen region for the same values of $\ba$), whereas for the other collapsed region with positive $\bs$, $\langle\ma\rangle$ is large.

The fluctuations $\sigma^2(\ma)$ and $\sigma^2(\ms)$, also shown in Figure \ref{s_as}, confirm this scenario. The fluctuations of $\ma$ sharply peak upon increasing  $\ba$. This peak is stronger for positive values of $\bs$, where it is matched by a peak in the fluctuations of $\ms$. For negative values, it is weaker, and there is no matching peak in the fluctuations of $\ms$. This indicates that there are two different transitions from the swollen region to the collapsed region. Moreover, within the collapsed region we find a sharp peak in the fluctuations in $\ms$ upon increasing $\bs$, which is not matched by a peak in the fluctuations of $\ma$.

\begin{figure}[ht!]
\centering
\includegraphics[width=0.45\textwidth]{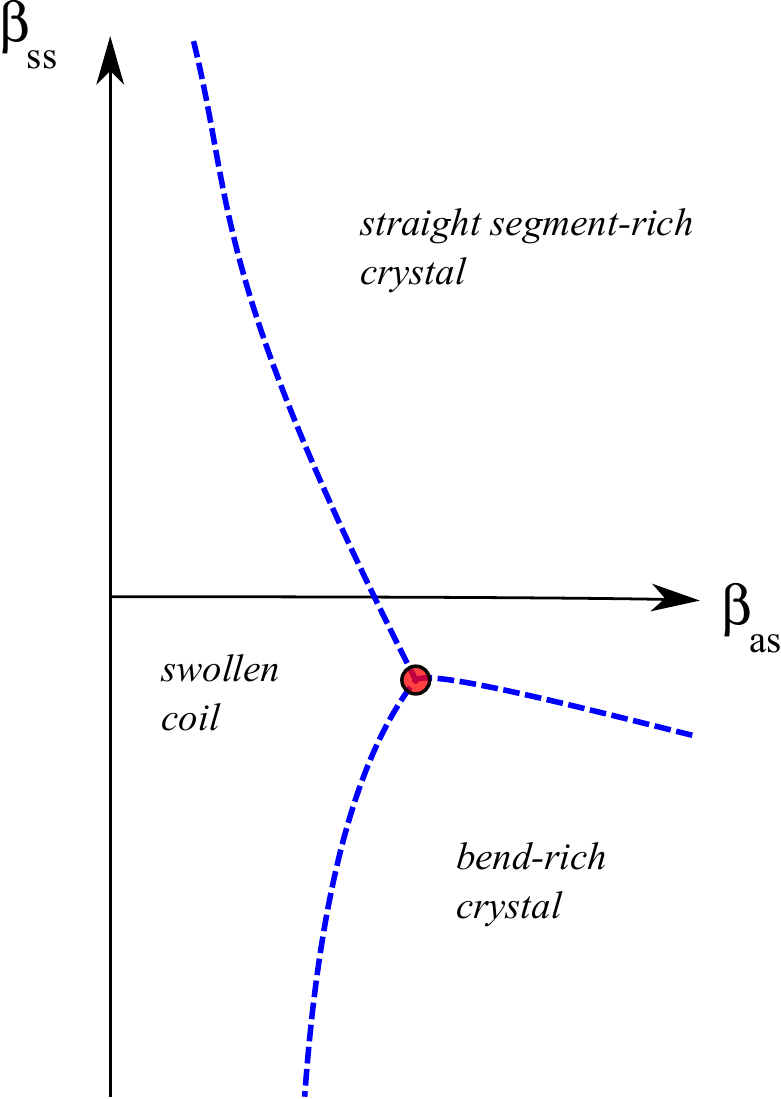}
\caption{\label{pd}
A sketch of the conjectured phase diagram, showing the swollen globular phase and the two crystalline phases. The three phase transition lines (dashed blue) are all first-order and meet in a common critical point (red).
}
\end{figure}

We thus have three clearly defined regions in the $\ba$,$\bs$-plane, delineated by rather sharp peaks in fluctuations. The presence of these fluctuations leads us to identify these regions with thermodynamic phases, so that we have a swollen globular phase, and two collapsed phases, one of which is rich in straight segments, and one of which is rich in bends. The conjectured phase diagram is shown in Figure \ref{pd}. This phase diagram also indicates that the transition lines are first-order, and that the collapsed phases have in fact a crystalline structure. We will provide evidence for this in the next two sections.

\subsection{Phase Transitions}

\begin{figure}[ht!]%
  \begin{center}
    \includegraphics[width=0.45\textwidth]{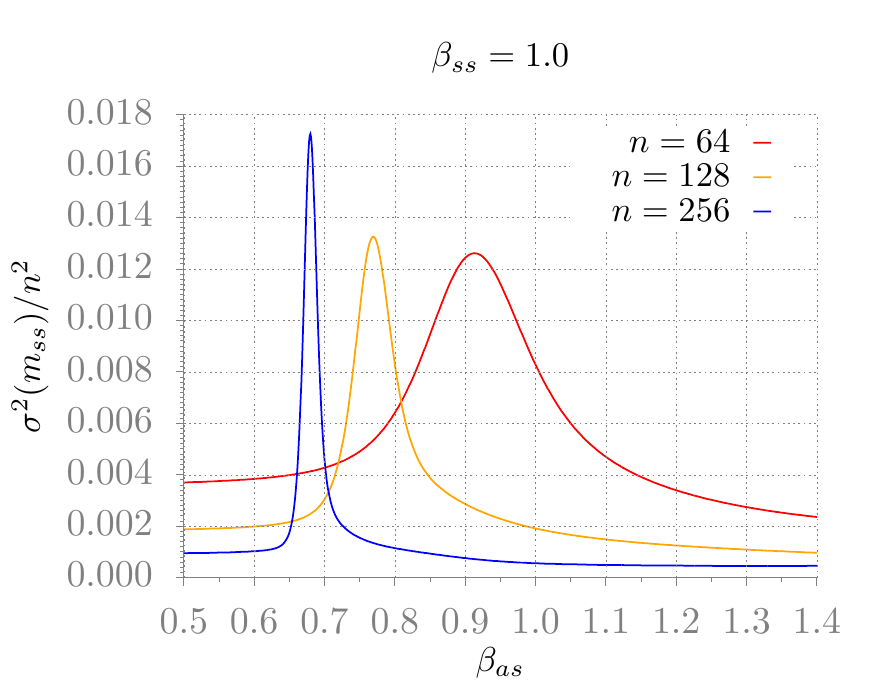} \hfill
    \includegraphics[width=0.45\textwidth]{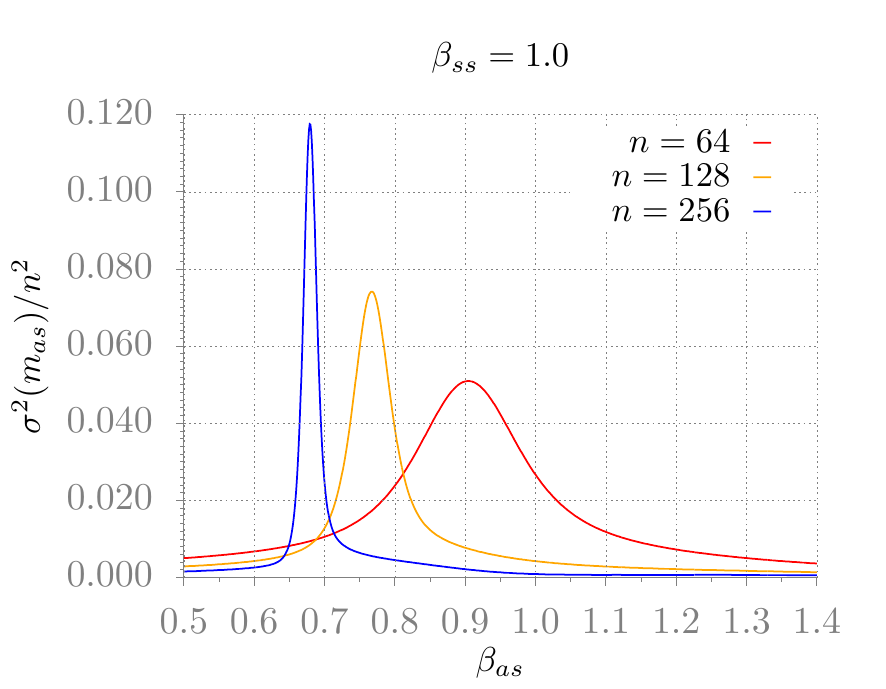} \\
    \includegraphics[width=0.45\textwidth]{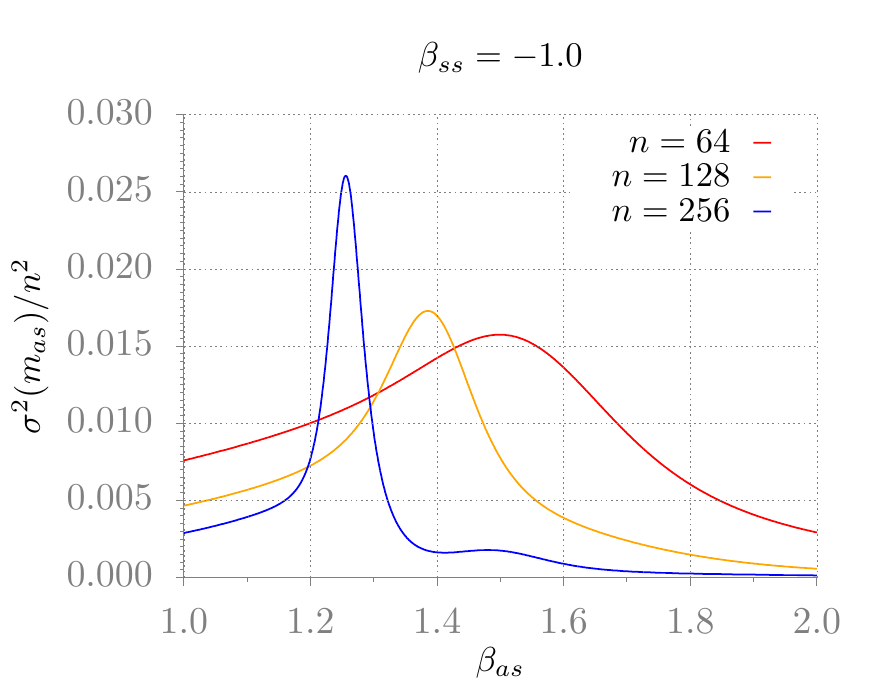} \hfill
    \includegraphics[width=0.45\textwidth]{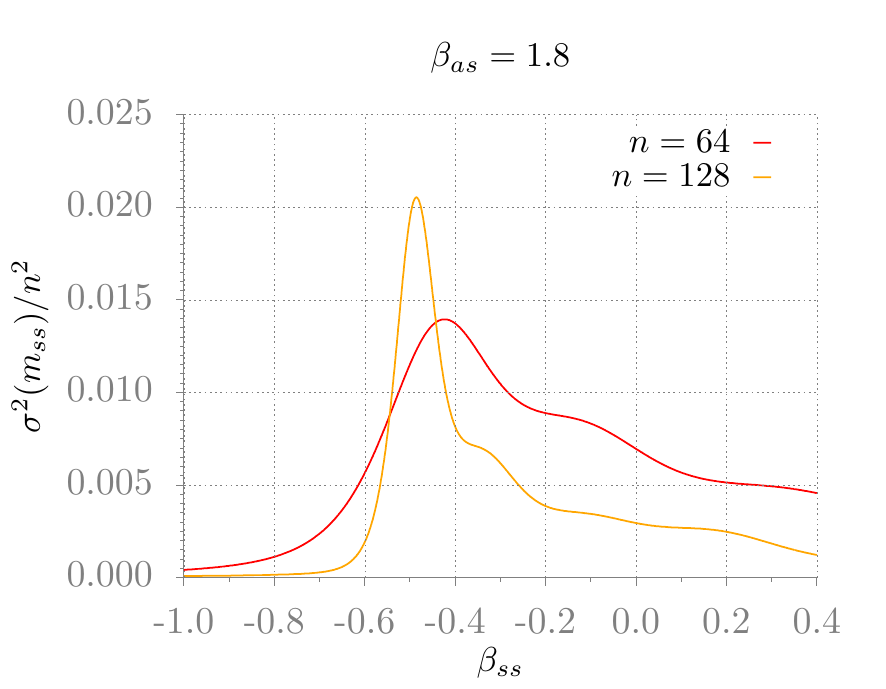} 
  \end{center}
  \caption{\label{fluct} Scaled fluctuations in $\ms$ and $\ma$ on the line $\bs=1.0$ (top left and right, respectively), scaled fluctuations in $\ma$ on the line $\bs=1.0$ (bottom left) and scaled fluctuations in $\ms$ on the line $\ba=1.8$ (bottom right)}
\end{figure}

\begin{figure}[b]
  \begin{center}
     \includegraphics[width=0.45\textwidth]{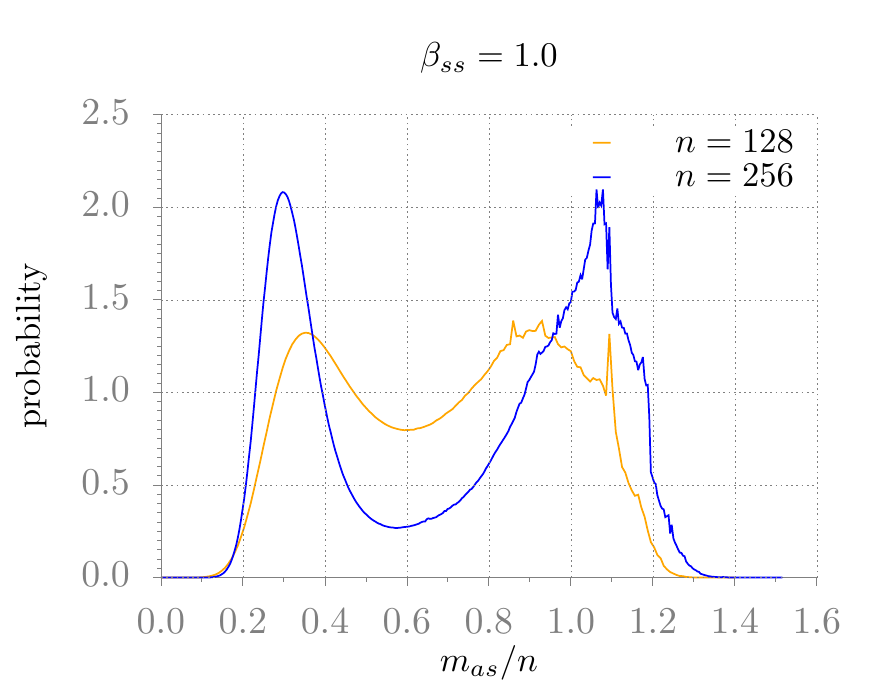}  \hfill
     \includegraphics[width=0.45\textwidth]{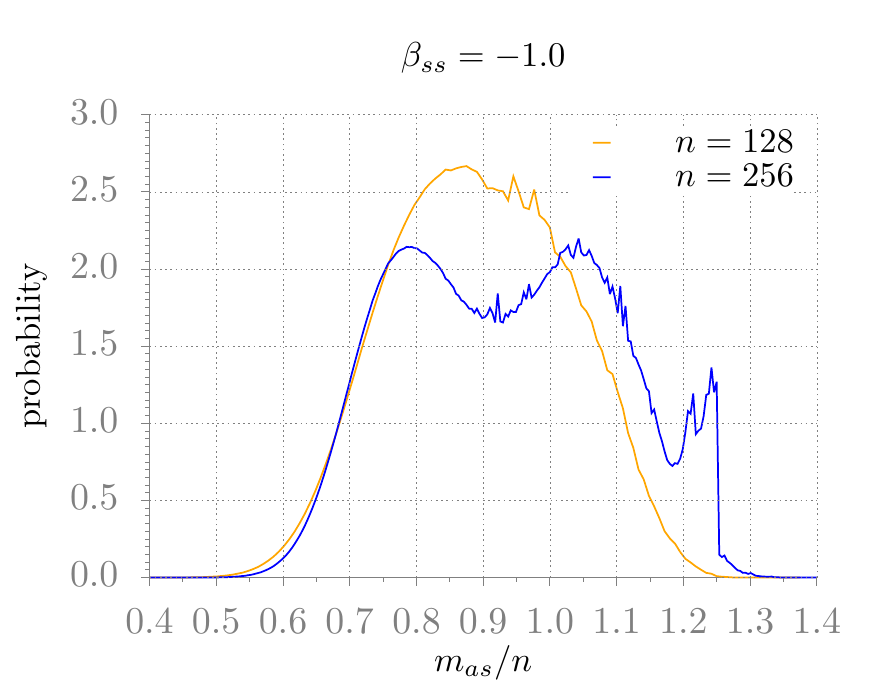}\\
     \includegraphics[width=0.45\textwidth]{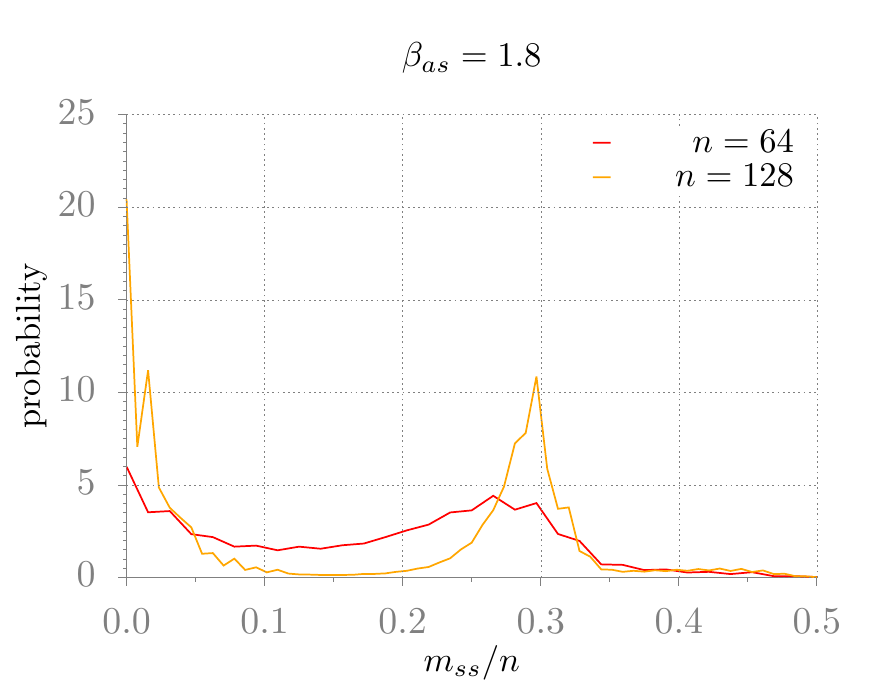} 
  \end{center}
  \caption{\label{hist}The top two figures show the scaled distributions of $\ma$ for $\bs=1.0$ and $\bs=-1.0$ (top left and right, respectively), for walk lengths $n=128$ and $n=256$, with the values of $\ba$ chosen such that
  the fluctuations in $\ma$ were maximal (for $\bs=1.0$, $\ba=0.770$ and $0.681$ for $n=128$ and $256$, respectively, and for $\bs=-1.0$, $\ba=1.382$ and $1.254$ for $n=128$ and $n=256$, respectively. 
  The bottom figure shows the scaled distributions of $\ms$ for $\ba=-1.8$ for walk lengths $n=64$ and $n=128$, with the values of $\bs$ chosen such that the fluctuations in $\ms$ were maximal ($\bs=-0.420$ and $-0.487$ for $n=64$ and $n=128$, respectively).
  }
\end{figure}

In order to study the order of the three transitions, we consider the behaviour of the system on the straight lines $\bs=1.0$, $\bs=-1.0$, and $\ba=1.8$, which have been chosen to cut across the three different phase transition lines. In Figure \ref{fluct} we show the scaled fluctuations $\sigma^2(\ma)/n^2$ and $\sigma^2(\ms)/n^2$. The scaling has been chosen such that the peak height should tend to a constant if the transition is first-order. Clearly all three lines show strong transitions, and to clarify whether they are really first-order we consider the distribution of the appropriate microcanonical parameters.

In Figure \ref{hist} we show the distribution of $\ma$ at the peak of its fluctuation in $\ba$ along the lines $\bs=1.0$ and $\bs=-1.0$, and the distribution of $\ms$ at the peak of its fluctuation in $\bs$ along the line $\ba=1.8$.
The distribution along $\bs=1.0$ is clearly bimodal with the distance between the double peaks slightly widening as $n$ changes from $128$ to $256$. This confirms the first-order nature of that transition, and also explains why the fluctuation peak grows super-linearly in Figure \ref{fluct}. Along the line $\bs=-1.0$ we also find an emerging double peak, albeit developing at much longer lengths. Finally, along the line $\ba=1.8$ we find a bimodal distribution with one peak located very closely to $\ms=0$. Hence, we conclude that the transitions between all three phases are first-order. 

\subsection{Nature of the Collapsed Phases}

\begin{figure}[ht!]
  \begin{center}
     \includegraphics[width=0.45\textwidth]{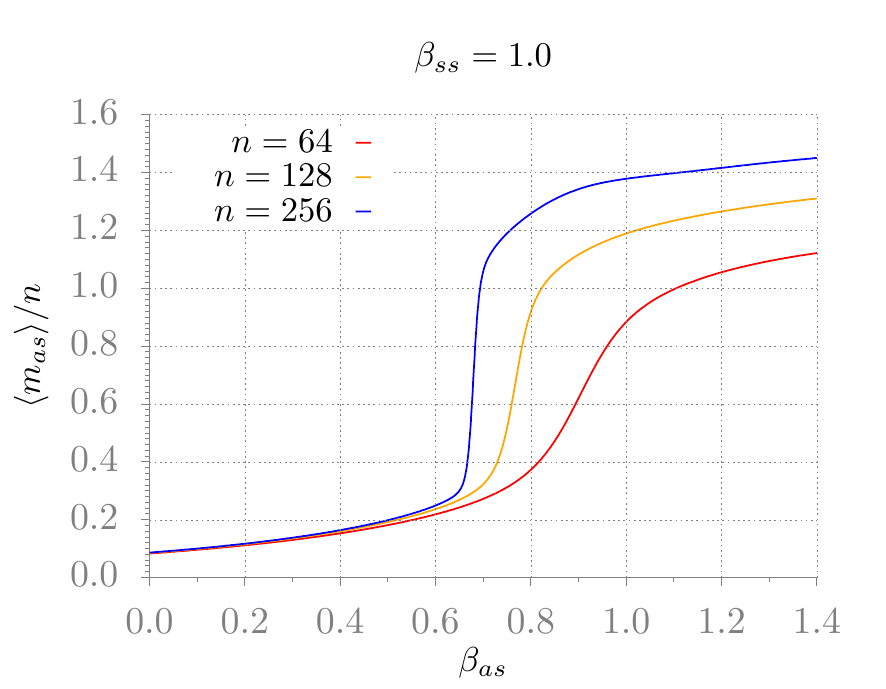} 
     \includegraphics[width=0.45\textwidth]{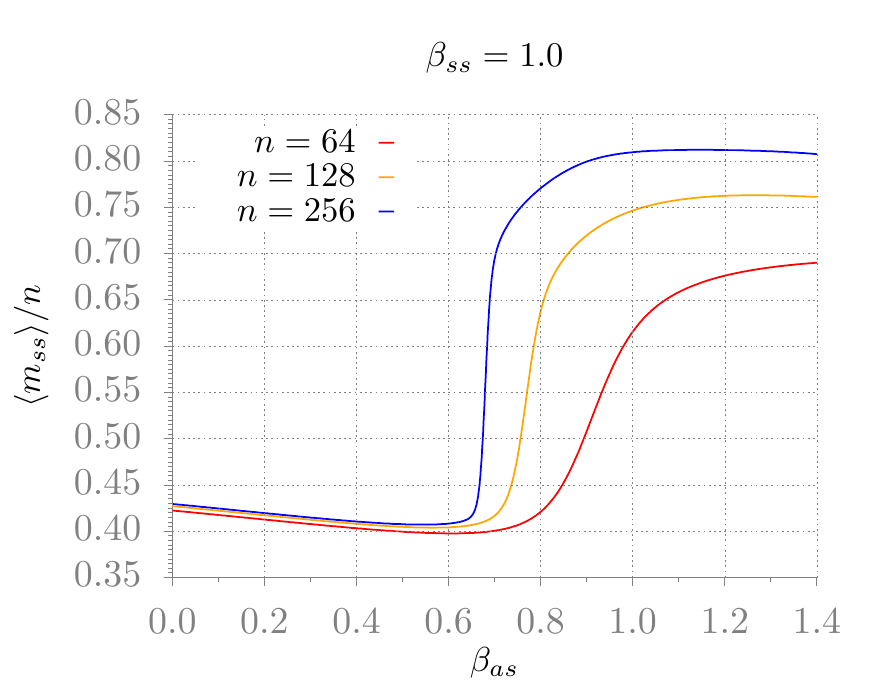}  \\  
     \includegraphics[width=0.45\textwidth]{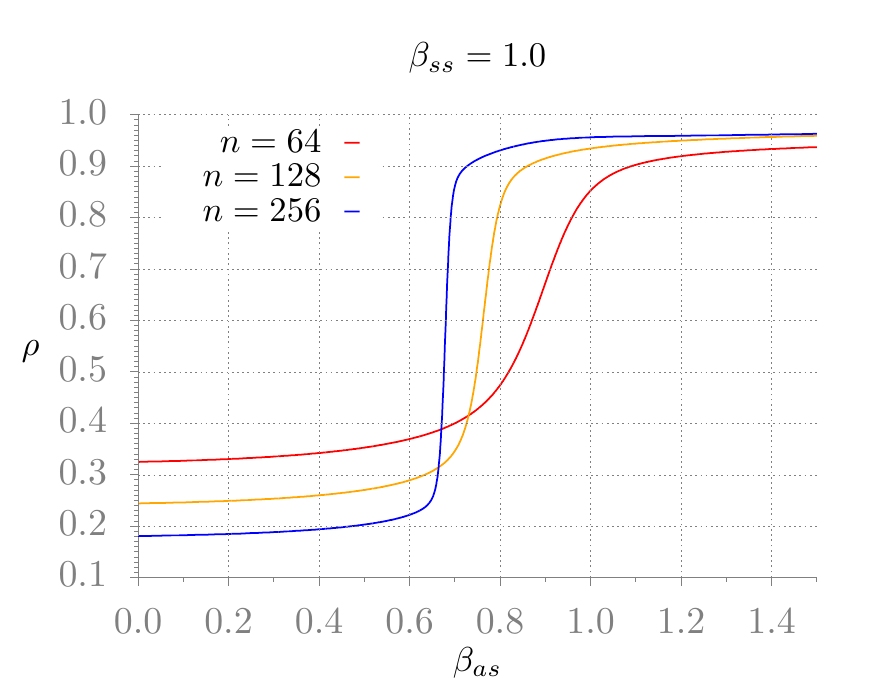} 
  \end{center}
  \caption{\label{bs1} Line $\bs=1.0$. The scaled number of contacts $\langle\ma\rangle$ and $\langle\ms\rangle$ and anisotropy
  parameter as a function of $\ba$ for constant $\bs=1.0$ for three system sizes $n=64,\;128$ and $256$.} 
\end{figure}

\begin{figure}[ht!]
  \begin{center}
     \includegraphics[width=0.45\textwidth]{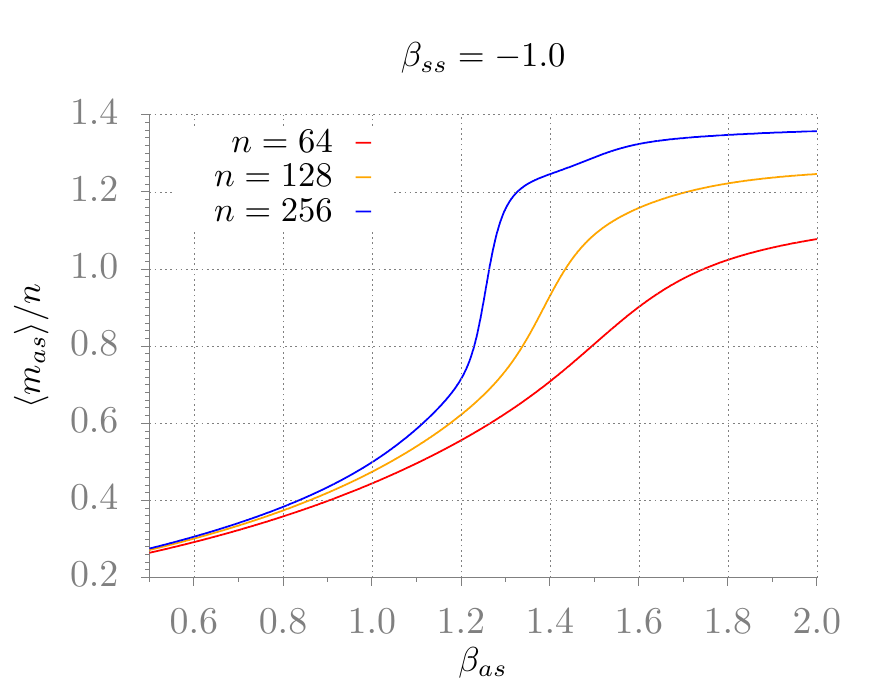}
     \includegraphics[width=0.45\textwidth]{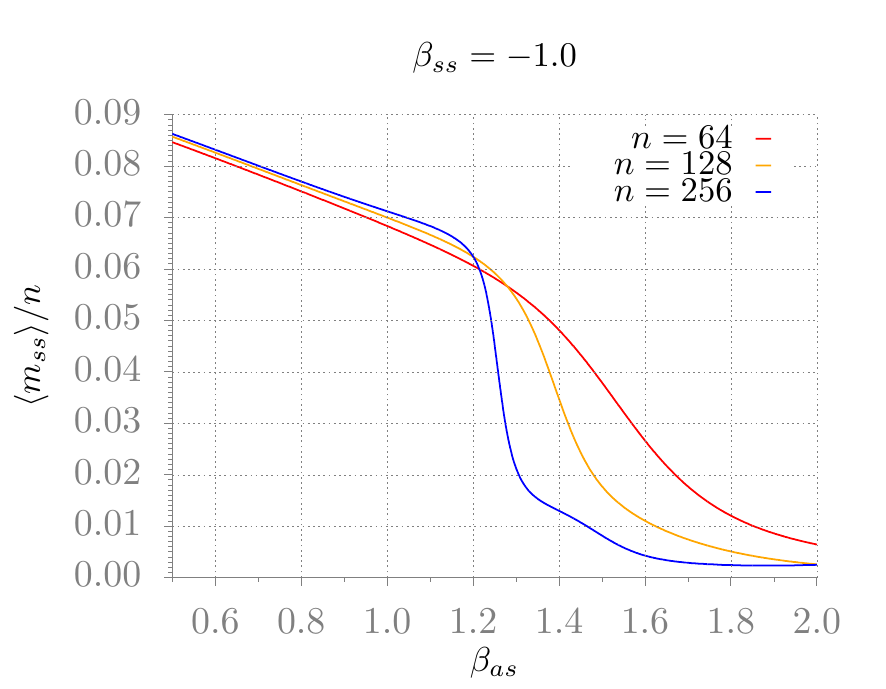}\\
     \includegraphics[width=0.45\textwidth]{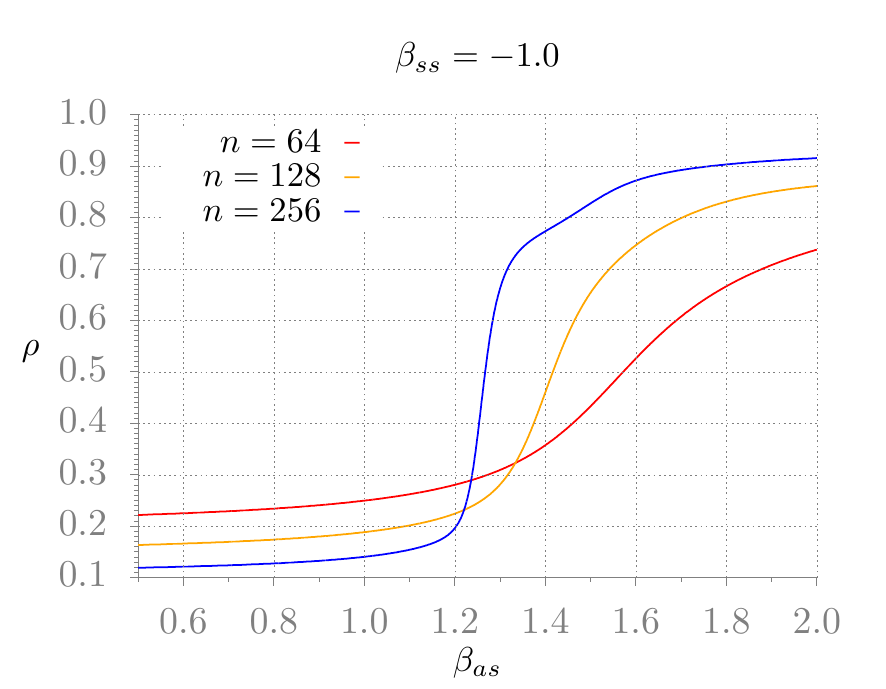}
  \end{center}
  \caption{\label{bs-1} Line $\bs=-1.0$.  The scaled number of contacts $\langle\ma\rangle$ and $\langle\ms\rangle$ and anisotropy
  parameter as a function of $\ba$ for constant $\bs=-1.0$ for three system sizes $n=64,\;128$ and $256$.} 
\end{figure}

\begin{figure}[ht!]
  \begin{center}
     \includegraphics[width=0.45\textwidth]{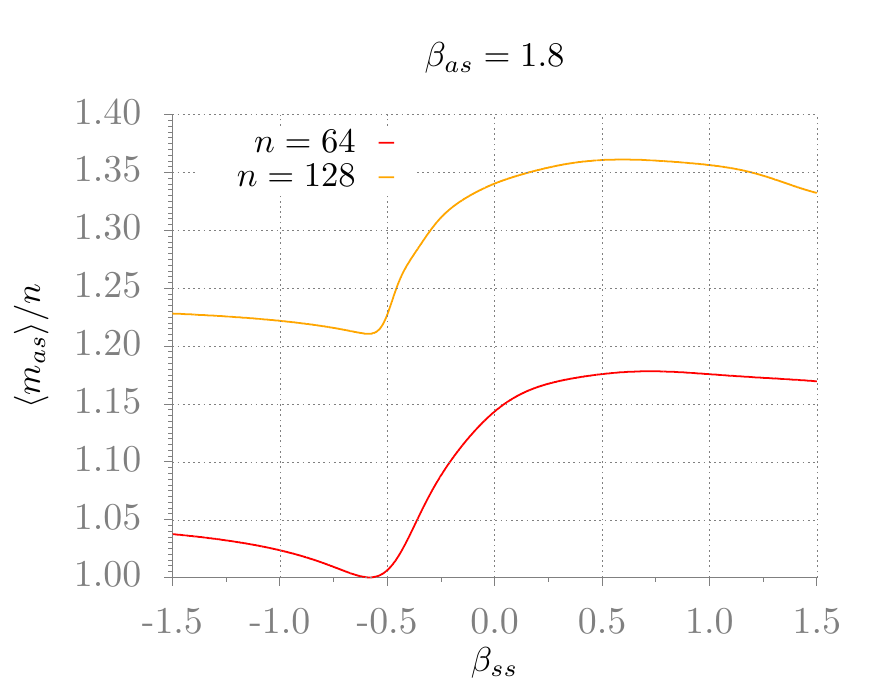}
     \includegraphics[width=0.45\textwidth]{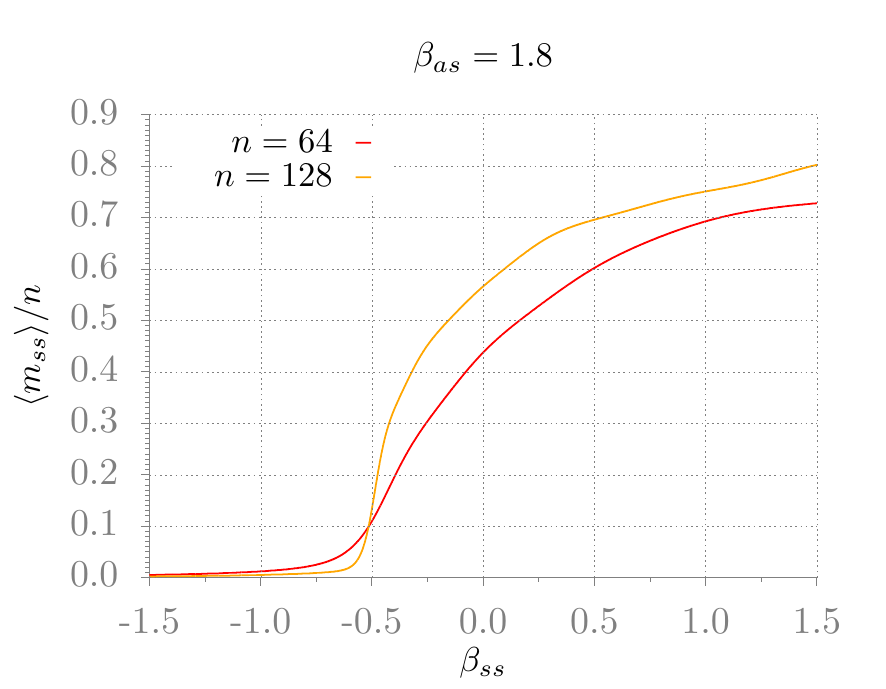}\\
     \includegraphics[width=0.45\textwidth]{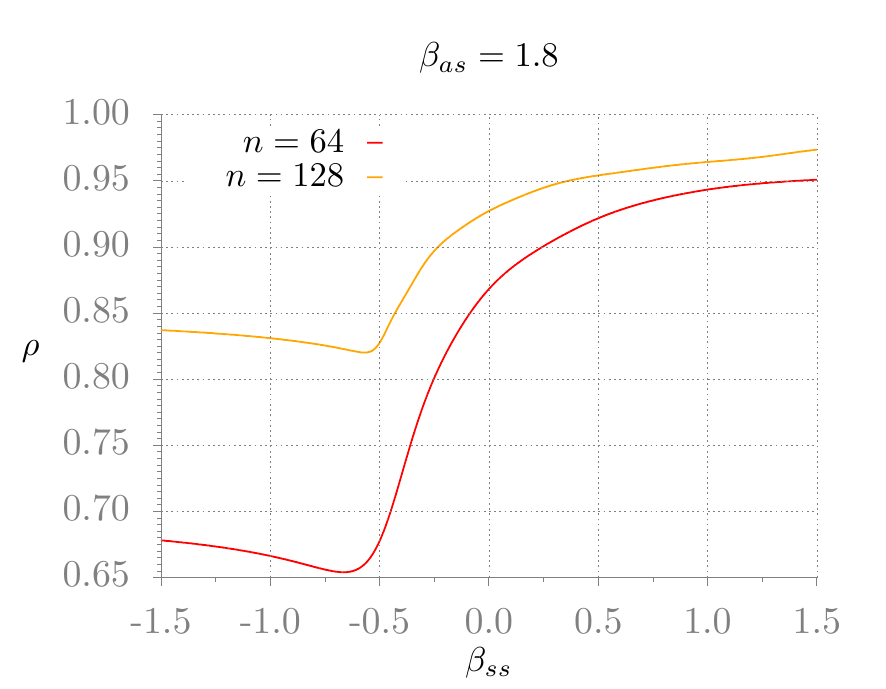}
  \end{center}
  \caption{\label{ba} Line $\ba=1.8$.  The scaled number of contacts $\langle\ma\rangle$ and $\langle\ms\rangle$ and anisotropy
  parameter as a function of $\bs$ for constant $\ba=1.8$ for two system sizes $n=64$ and $128$.} 
\end{figure}

We now turn to the investigation of the two collapsed phases. As in \cite{krawczyk2010a-:a}, it will be helpful to consider an anisotropy parameter. In three dimensions, denoting the number of bonds parallel to the $x$-, $y$-, 
and $z$-axes by $n_x$, $n_y$, and $n_z$, respectively, we define
\begin{equation}
\rho=1.0-\frac{\min(n_x,n_y,n_z)}{\max(n_x,n_y,n_z)}
\end{equation}
to be the anisotropy parameter.
In a system without orientational order, this quantity tends to zero
as the system size increases.  A non-zero limiting value less than one
of this quantity indicates weak orientational order with
$n_{min}\propto n_{max}$, while a limiting value of one indicates
strong orientational order, where $n_{max}\gg n_{min}$.

We first consider the change of the anisotropy parameter as the phase boundaries are crossed along the same three lines investigated 
earlier. This is shown in Figure \ref{bs1} for $\bs=1.0$, in Figure \ref{bs-1} for $\bs=-1.0$, and in Figure \ref{ba} for $\ba=1.8$, together with the corresponding changes of the density of interacting segments $\langle\ma\rangle$ and straight segments $\langle\ms\rangle$.

In addition to the change in $\langle\ma\rangle$ and $\langle\ms\rangle$ described above, one sees that the anisotropy parameter decreases in the swollen phase, but jumps to a significantly larger value in the collapsed phases. There is a small difference in the value of the anisotropy parameter in both collapsed phases, but the change of $\rho$ as the length of the walk increases is such that $\rho$ might in fact tend to the same limiting value of one in the thermodynamic limit.

\begin{figure}[b]
  \begin{center}
     \includegraphics[width=0.45\textwidth]{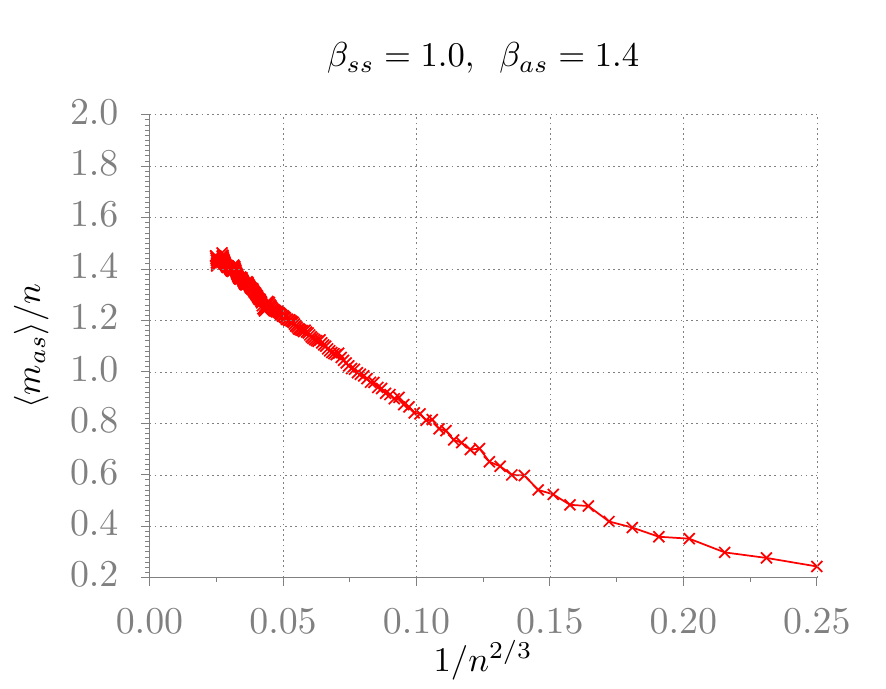}  \hfill 
     \includegraphics[width=0.45\textwidth]{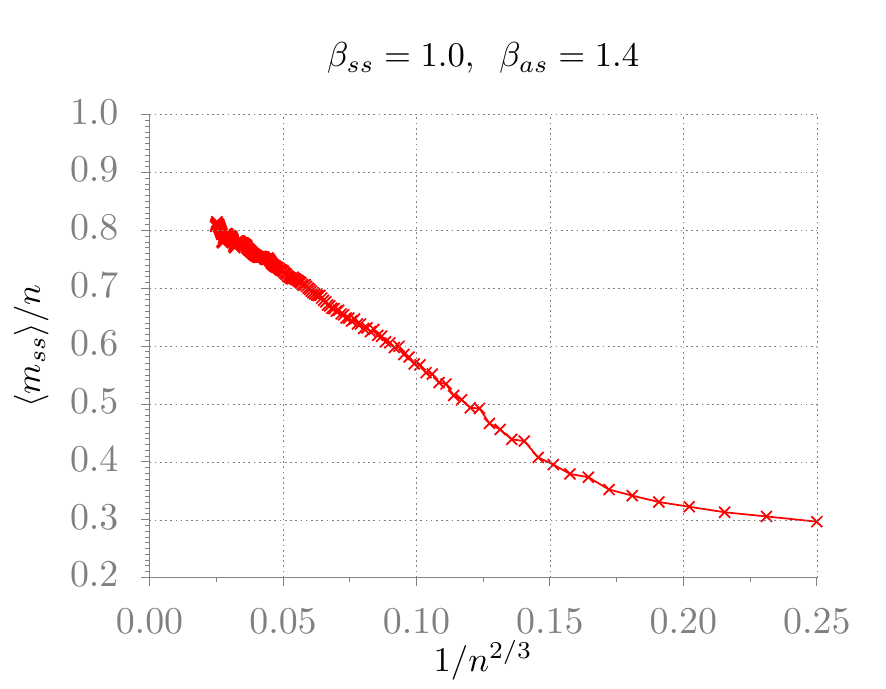}  \\
     \includegraphics[width=0.45\textwidth]{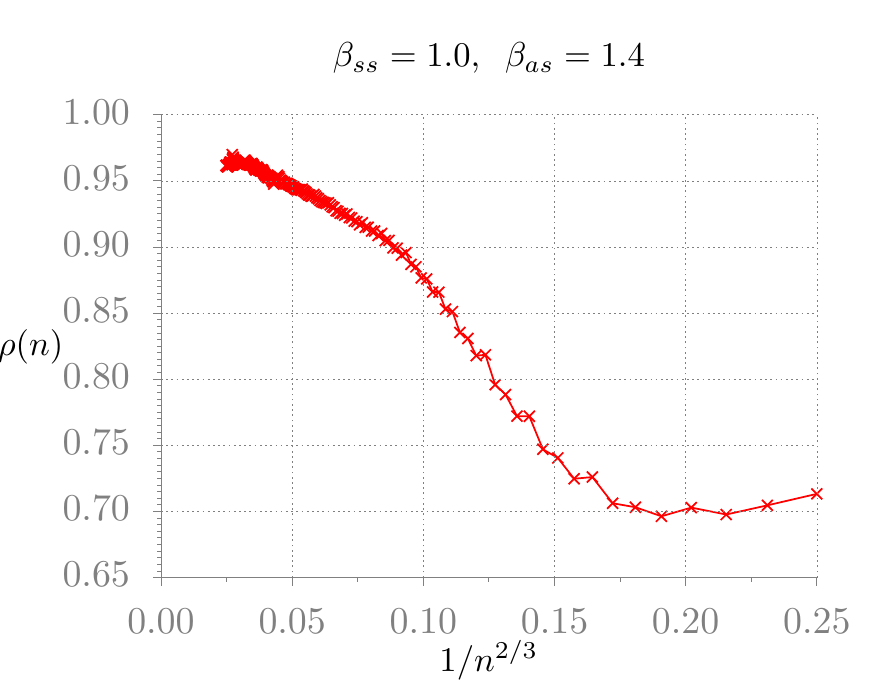}
  \end{center}
  \caption{\label{point1}Number of contacts $\langle\ms\rangle/n$  and $\langle\ma\rangle/n$ and anisotropy parameter, $\rho$, as function of the system size at $\ba=1.4$ and $\bs=1.0$. Importantly, the anisotropy parameter, $\rho$, can be seen to converge to $\rho=1$ for infinite $n$ within error.  The thermodynamic average of straight segments converges to a value near $0.9$ within error.
  }
\end{figure}

\begin{figure}[b]
  \begin{center}
     \includegraphics[width=0.45\textwidth]{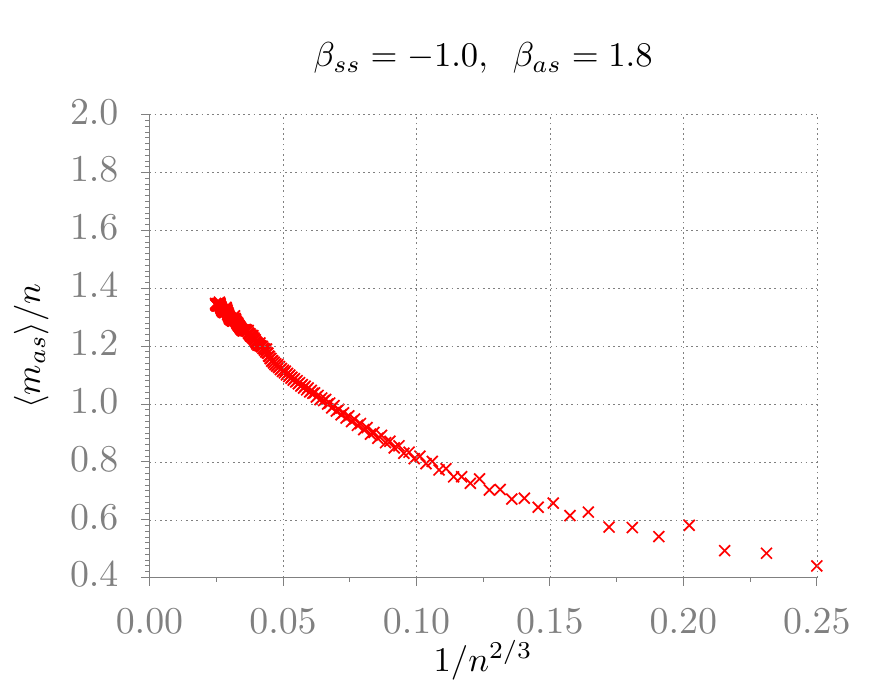}  \hfill 
     \includegraphics[width=0.45\textwidth]{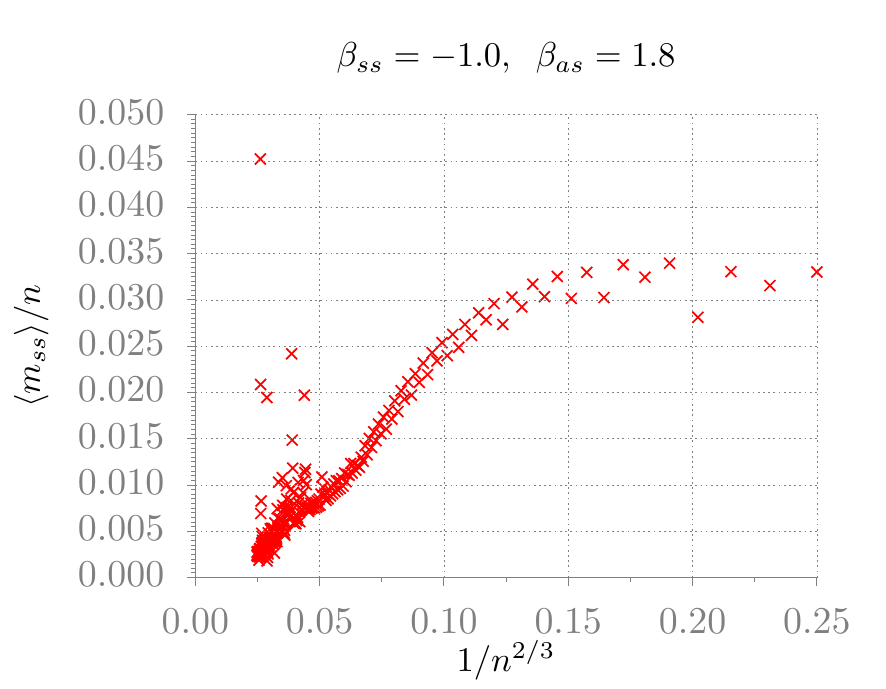}  \\ 
     \includegraphics[width=0.45\textwidth]{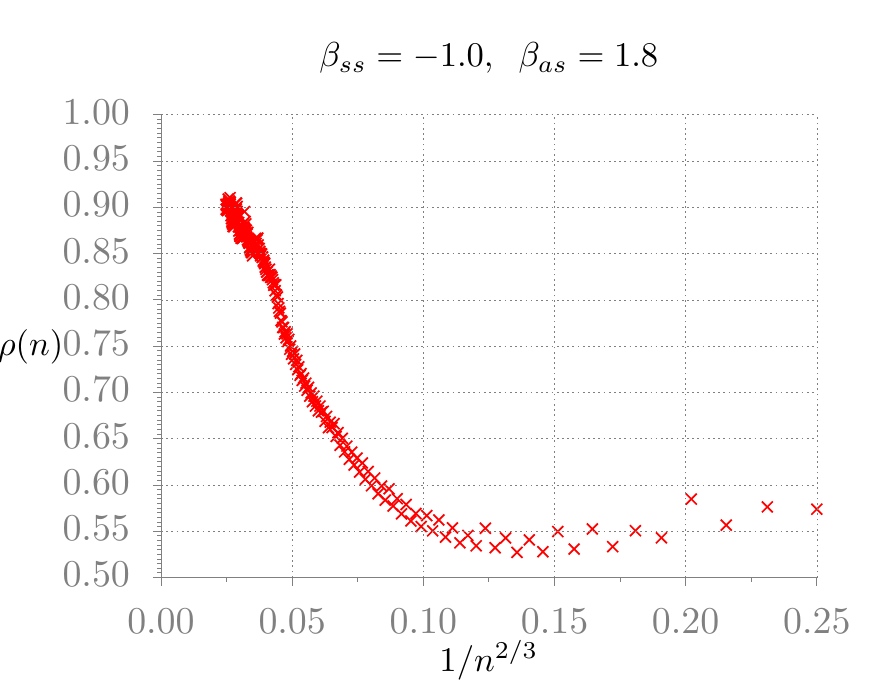} 
  \end{center}
  \caption{\label{point2}Number of contacts $\langle\ms\rangle/n$  and $\langle\ma\rangle/n$ and anisotropy parameter, $\rho$, as function of the system size at $\ba=1.8$ and $\bs=-1.0$.  Importantly, the anisotropy parameter, $\rho$, can be seen to converge to $\rho=1$ for infinite $n$ within error and the thermodynamic average of straight segments converges to $\lim_{n\rightarrow\infty}\langle\ms\rangle/n =0$ within error.
  }
\end{figure}

To test this, we now display the finite-size scaling behaviour at two points in the different collapsed phases. The values $\langle\ms\rangle/n$, $\langle\ma\rangle/n$ and $\rho$ are shown
as functions of $n^{-2/3}$ for $\ba=1.4$ and $\bs=1.0$ in Figure \ref{point1} and for $\ba=1.8$ and $\bs=-1.0$ in Figure \ref{point2}. If the corrections to scaling are due to surface effects,
one should find asymptotic straight lines as $n^{-2/3}$ tends to zero. 

Both figures confirm the presence of $n^{-2/3}$-corrections, indicating a well-developed surface of the collapsed walk. The anisotropy parameter tends to one in both phases in the thermodynamic limit,
indicating crystalline order. However, in one phase the density of straight segments tends to a limiting value close to one, whereas in the other phase this quantity tends to zero, showing clear evidence
of two structurally very different crystalline structures, each of which has a different limiting density of attracting segments. 

\begin{figure}[b]
  \begin{center}
     \includegraphics[width=0.45\textwidth]{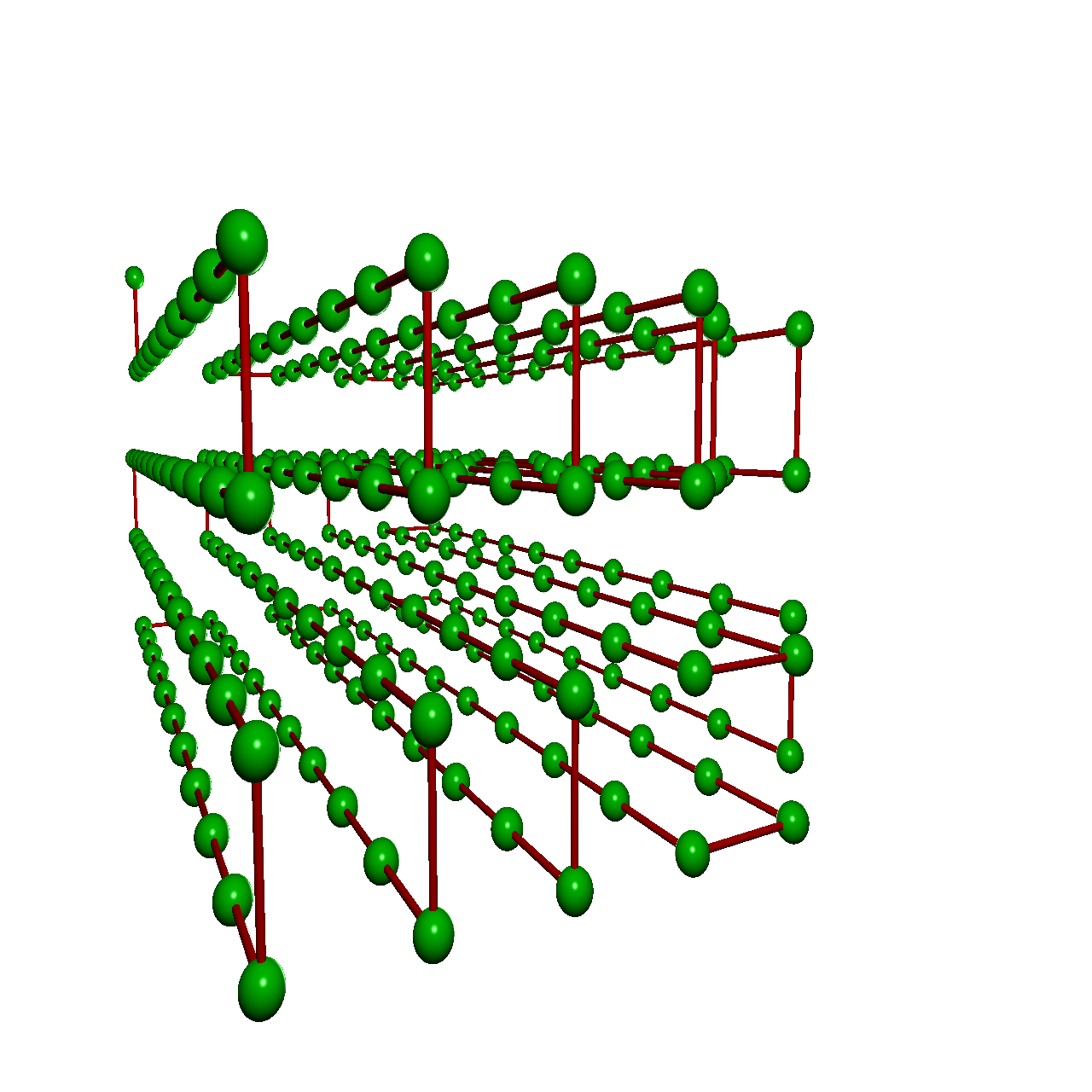}  
     \includegraphics[width=0.45\textwidth]{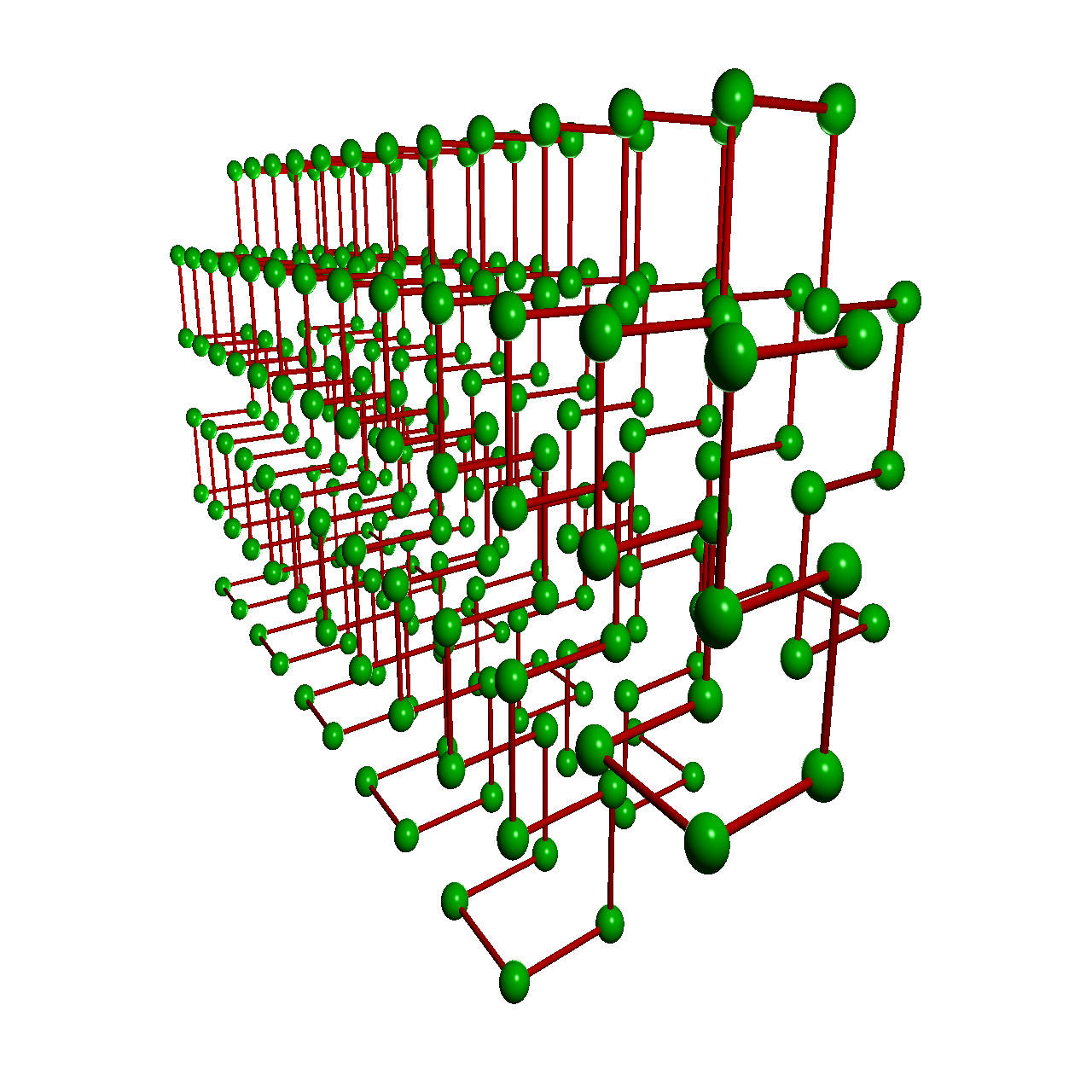}   

  \end{center}
  \caption{\label{typical} Typical configurations of length 128 at the parameter values $(\ba,\bs)=(1.4,1.0)$ (left) and $(\ba,\bs)=(1.8,-1.0)$ (right). The configuration on the left is an array of long straight lines of monomers while the configuration on the right contains mainly bends.
  }
\end{figure}

This scenario is confirmed by Figure \ref{typical}, which shows typical configurations at these two parameter values: in the straight-rich phase,  $(\ba,\bs)=(1.4,1.0)$, the typical configurations consists of straight lines of monomers while in the bend rich phase, $(\ba,\bs)=(1.8,-1.0)$, the typical configuration can be described as parallel sheets of two-dimensional configurations consisting only of bends. The limiting value of $\langle\ma\rangle/n$ is equal to $2$ in the former phase, and equal to $7/4$ in the latter phase, consistent with the numerical data shown in Figure \ref{point1} and Figure \ref{point2}.

In this article we have discussed a three-dimensional  semi-flexible AS model. We have characterised one swollen
and two collapsed phases, and transition between them. In contrast to the two dimensional version of this model,
both collapsed phases are anisotropic, and none of them resemble the disordered liquid drop.

\section*{Acknowledgements}

Financial support from the Australian Research Council via its support
for the Centre of Excellence for Mathematics and Statistics of Complex
Systems and via its Discovery program is gratefully acknowledged by the authors. A L Owczarek thanks the
School of Mathematical Sciences, Queen Mary, University of London for
hospitality. This research was supported in part by Polish Grid Infrastructure.

\end{document}